\documentclass[aps,preprint,amsmath]{revtex4}

\def\al{\alpha}

\def\ga{\gamma}

\def\ze{\zeta}

\def\ps{\psi}

\def\De{\Delta}

\def\cE{{\cal E}}
\def\cl{{\cal L}}

\def\mn{{\mu\nu}}

\def\half{{\textstyle{1\over 2}}}
\def\quar{{\textstyle{1\over 4}}}
\def\frac#1#2{{\textstyle{{#1}\over {#2}}}}

\def\vev#1{\langle {#1}\rangle}

\def\lsim{\mathrel{\rlap{\lower4pt\hbox{\hskip1pt$\sim$}}
    \raise1pt\hbox{$<$}}}
\def\gsim{\mathrel{\rlap{\lower4pt\hbox{\hskip1pt$\sim$}}
    \raise1pt\hbox{$>$}}}
\def\sqr#1#2{{\vcenter{\vbox{\hrule height.#2pt
         \hbox{\vrule width.#2pt height#1pt \kern#1pt
         \vrule width.#2pt}
         \hrule height.#2pt}}}}

\def\pt#1{\phantom{#1}}
\def\ol#1{\overline{#1}}

\def\etal{{\it et al.}}

\def\hul#1#2{h^{#1}_{{\pt{#1}}#2}}

\def\ab{\overline{a}}
\def\atw{\widetilde{a}}

\def\mt{m^{\rm T}}
\def\ms{m^{\rm S}}

\def\oaut{{\ol a}^{\rm T}}
\def\oaus{{\ol a}^{\rm S}}

\def\lrpartial{\raise 1pt\hbox{$\stackrel\leftrightarrow\partial$}}

\newcommand{\beq}{\begin{equation}}
\newcommand{\eeq}{\end{equation}}
\newcommand{\bea}{\begin{eqnarray}}
\newcommand{\eea}{\end{eqnarray}}
\newcommand{\rf}[1]{(\ref{#1})}

\begin{document}

\title{Prospects for Large Relativity Violations in\\
Matter-Gravity Couplings}

\author{V.\ Alan Kosteleck\'y and Jay D. Tasson}
\affiliation{Physics Department, Indiana University,
Bloomington, IN 47405, U.S.A.} 

\date{IUHET 519, October 2008}

\begin{abstract}
Deviations from relativity are tightly constrained
by numerous experiments.
A class of unmeasured and potentially large violations is presented
that can be tested in the laboratory only via weak gravity couplings.
Specialized highly sensitive experiments
could achieve measurements of the corresponding effects.
A single constraint of $1 \times 10^{-11}$ GeV
is extracted on one combination
of the 12 possible effects in ordinary matter.
Estimates are provided for attainable sensitivities
in existing and future experiments.
\end{abstract}

\maketitle

Einstein's theories of special and general relativity
form the underpinning of our best existing
description of nature at the fundamental level.
The key idea behind relativity 
is the notion of Lorentz symmetry:
the invariance of the laws of physics 
under rotations and boosts of the system.
Experimental testing of these ideas 
has achieved impressive sensitivities
to hypothetical tiny deviations from Lorentz symmetry
in special relativity
\cite{tables},
with several tests using matter and light
now well below parts in $10^{30}$.
For general relativity,
the situation is more challenging 
because gravity is a weak force on small scales
\cite{will}.
Recently,
sensitive new constraints on violations of Lorentz symmetry
in general relativity have been obtained 
\cite{gravexpt}.

Given the remarkable experimental sensitivities attained 
and the breadth of the studies,
a question of immediate interest is whether 
any types of comparatively large relativity violations
could have evaded detection to date. 
Here,
we show the answer is affirmative.
We demonstrate the existence of a type of Lorentz violation 
that is natural,
challenging to observe in tests of special relativity,
and directly detectable in laboratory experiments
only when suppressed by weak gravitational effects.
A general framework is given for studying 
this hidden type of violation,
a constraint is obtained on one combination 
of the 12 possible effects in ordinary matter,
and prospects for future measurements 
in specialized experiments are examined.

An arbitrary Lorentz violation represents
a violation of rotation or boost symmetry
and hence can be characterized
via a nonzero vector or tensor quantity
in the vacuum
\cite{kp95}.
The specific violation of interest here 
involves an observer 4-vector $a_\mu$
that couples to a fermion field $\ps$
as a term $\cl_a = -a_\mu\ol\ps\ga^\mu\ps$ in the Lagrange density.
This coupling is comparatively simple and theoretically natural.
As it is quadratic in $\ps$,
it modifies the fermion dispersion relation.
For example,
for a constant $a_\mu$ in Minkowski spacetime,
a free fermion of mass $m$, energy $E$, 
and momentum $\vec p$
acquires the dispersion relation
\cite{ck}
\beq
(E-a_0)^2 = m^2 c^4 + (\vec p - \vec a)^2c^2.
\eeq 
A fermion at rest can be shown to have 
$\vec p = \vec a$
and would therefore satisfy a modification
of Einstein's famous equation relating matter and energy: 
$E = mc^2 + a_0$.
More generally, 
the coupling $a_\mu$
can depend on the species $w$ of fermion
and is denoted $a^w_\mu$.

Although at first glance a nonzero $a_\mu$ appears to be
a substantial modification of known physics,
in fact it is challenging to observe experimentally.
In Minkowski spacetime,
where gravity is irrelevant,
a constant $a_\mu$
could be arbitrarily large 
because the coupling $\cl_a$ is unobservable
\cite{ck}
in experiments with a single fermion flavor.
Under these circumstances,
$a_\mu$ can be absorbed by a phase shift of the fermion field,
which is a canonical transformation
reflecting the inherent ambiguity of measuring 
absolute values of energy and momentum.
Only the difference $\De a_\mu$ between two fermion flavors
is potentially observable,
and even if nonzero this requires special experiments 
involving flavor-changing fermions
such as neutral-meson oscillations 
\cite{mesons}
or neutrino oscillations 
\cite{nu}.

In weak gravitational fields
such as those in our solar system,
the effects of gravity can be understood
as a perturbation $h_{\mu\nu}$ 
in a background Minkowski spacetime.
A constant $a_\mu$
could still be absorbed by a phase shift
and would remain strictly undetectable.
However,
$a_\mu$ cannot be constant generically
because it must be compatible with the geometrical structure
of gravity
\cite{akgrav}.
In essence,
the interaction of $a_\mu$ with the gravitational field 
ensures that $a_\mu$ varies with spacetime position,
and this implies only a single component 
of $a_\mu$ can be absorbed.

It is convenient to separate $a_\mu$ 
into a constant piece $\ol a_\mu$
and a fluctuation piece $\widetilde a_\mu$ 
arising from the gravitational interaction:
$a_\mu = \ol a_\mu + \widetilde a_\mu$.
In the vacuum,
the fluctuation $\widetilde a_\mu$
is tightly constrained by the requirements
of geometric compatibility 
and coordinate independence of the physics. 
These give rise to $\widetilde a_\mu$ of the form
\beq
\atw_\mu = 
\half \al h_\mn \ab^\nu - \quar \al \ab_\mu \hul{\nu}{\nu} 
\label{aftw}
\eeq
in harmonic coordinates,
where the constant $\al$ is determined by the
strength of the coupling of $a_\mu$ to gravity.
The gravitational field $h_\mn$ itself also acquires
a correction $\widetilde h_\mn$,
given at leading order by
$\widetilde h_{00} = 2 \al \ab_0 h_{00}/m$,
which avoids self-accelerations
and ensures that Newton's third law holds
between gravitating bodies.

The key point is that,
although a nonzero constant $\ol a_\mu$ remains directly unobservable,
its existence can be indirectly established 
through the effects of the $\widetilde a_\mu$ coupling. 
In particular,
since fluctuations in $h_{\mu\nu}$ are tiny in the solar system,
the coefficient $\ol a_\mu$ can be 
enormous compared to other effects, 
while having evaded detection in all experimental tests
of relativity to date.
The validity of perturbation theory requires $\ol a_\mu$
to be less than the fermion mass $m$,
but this still leaves room for effects 
some $10^{30}$ times greater than the best existing constraints 
\cite{tables}
on other types of relativity violation.
Indeed,
the theoretically allowed values of $\ol a_\mu$
are large enough to obviate the Lorentz hierarchy problem
\cite{kp95},
since they could lie within a few orders of the fermion mass.
Radiative corrections involving two powers of $\ol a_\mu$ 
and a gravitational coupling
could in principle produce effects in nongravitational experiments
searching for other coefficients for Lorentz violation
such as 
an observer two-tensor $\ol c_{\mu\nu}$,
but even for large $\ol a_\mu$ 
the resulting signals would be far below current sensitivities. 
Evidently, 
the detection of $\ol a_\mu$ 
requires specialized gravitational experiments
of high sensitivity.

What kind of field theory can produce an $a_\mu$ coupling?
The geometric structure of gravity constrains
the violation of Lorentz symmetry to be spontaneous
rather than explicit
\cite{akgrav},
so the theory involves a Lorentz-tensor field 
that acquires a nonzero vacuum value.
Since $a_\mu$ has a single index,
the simplest choice is a vector field,
denoted $B_\mu$,
although other tensor fields can be considered
\cite{ks}. 
Vector theories with spontaneous breaking of Lorentz symmetry,
generically called bumblebee theories,
exist in many forms.
Here,
it suffices to suppose that the bumblebee field $B_\mu$
has a curvature coupling 
$\cl_B \supset \xi B^\mu B^\nu R_{\mu\nu}$ 
and a coupling to the fermion field
\cite{kl} 
$\cl_{\ps}\supset - \ze B_\mu \ol\ps \ga^\mu \ps$,
where $\xi$ and $\ze$ are coupling constants.
In this class of models,
the bumblebee vacuum value $\vev{B_\mu} = b_\mu$
produces a relativity violation of the $a_\mu$ type,
with the identification 
$\al = -4\xi$, $\ol a_\mu = \ze b_\mu$.

Spontaneous symmetry breaking
is accompanied by massless 
modes called Nambu-Goldstone (NG) modes,
which in the present context can be identified with
vacuum fluctuations $\cE_\mu$ of the bumblebee field
\cite{bkgrav}
or equivalently with the fluctuation
$\widetilde a_\mu = \ze \cE_\mu$.
In typical models,
the NG modes play the role of a long-range force.
They have previously been interpreted as the photon 
\cite{bkgrav}, 
the graviton 
\cite{kpgrav}, 
and a spin-dependent interaction
\cite{ah}.
Here, 
the NG modes play a different role:
mediating a spin-independent force between fermions,
with coupling constant $\ze$ 
also controlling Lorentz violation. 
New spin-independent forces are constrained by experiments
\cite{schlamminger},
which in this context limit the strength of $\ze$
but not the size of the Lorentz violation $\ol a_\mu$.

Numerous scenarios for $\ol a_\mu$ can be considered,
depending on properties of the coupling $\ze$
and the vacuum value $b_\mu$,
and there is a correspondingly wide variety 
of potentially observable signals. 
The coupling $\ze$ and hence the coefficient $\ol a_\mu$ 
may be flavor independent
or may depend on properties of the fermion.
For example,
it could be proportional to the fermion mass $m$,
in analogy with the usual Yukawa couplings.
Alternatively, 
it may depend on other quantum numbers
such as baryon number $B$, lepton number $L$,
or combinations of these such as 
the difference $B-L$ that is conserved in many
grand unified theories.
It could be proportional 
to the fermion charge $Q$,
as occurs in bumblebee electrodynamics
\cite{bkgrav}.
This further hides the Lorentz violation
because effects cancel in charge-neutral matter,
so observable signals in this case require
specialized experiments designed to study
the effects of gravity on charged matter,
such as electron interferometry
\cite{emepspace}.
Another scenario has effects
from $\ol a_\mu$ cancelling against
those from different unmeasured coefficients for Lorentz violation 
such as an observer two-tensor $\ol c_{\mu\nu}$,
so that signals in ordinary matter would be absent.
Since $\ol a_\mu$ violates CPT symmetry
while $\ol c_{\mu\nu}$ is invariant,
this cancellation implies an observable enhancement
in future gravitational experiments with antihydrogen
\cite{antih}
or antiparticles
\cite{muonium}.
 
The vacuum value $b_\mu$ and hence the coefficient $\ol a_\mu$ 
could be timelike, lightlike, or spacelike,
with different observable signals in each case.
Substantial differences between the magnitudes 
of components of $\ol a_\mu$ can be generated naturally.
For example,
if $\ol a_\mu$ is timelike,
then there exists an observer frame $O$
in which it is \it purely \rm timelike.
Provided the domain size is cosmological,
it may be appropriate to identify $O$ with the rest frame $U$
of the cosmic microwave background radiation.
In effect,
this aligns the Lorentz violation 
with the cosmological expansion,
thereby preserving isotropy
\cite{akgrav}.
However,
experiments are performed and reported locally in the solar system,
for which it is appropriate and conventional
to adopt a Sun-centered frame $S$ 
\cite{km}.
Since $S$ differs from $U$ by a boost,
in $S$ the spatial components $\ol a_J$ 
are nonzero but suppressed by a factor
of about 1000 relative to the temporal component $\ol a_T$.
As another example,
if $\ol a_\mu$ is spacelike instead,
then there exists an observer frame $O^\prime$
in which it is \it purely \rm spacelike.
If $O^\prime$ happens to coincide with $U$,
then in $S$ the temporal component $\ol a_T$ 
is nonzero but suppressed by a factor
of about 1000 relative to the spatial components $\ol a_J$.

To detect effects from $\ol a_\mu$,
the relevant experiments must be sensitive to gravity.
In a laboratory frame $L$,
it suffices to achieve sensitivity to modifications 
of the dominant local gravitational acceleration $g$.
The effects predicted by $\cl_a$ can be extracted
in the weak-gravity approximation
and at leading order in $\ol a_\mu$ and $h_{\mu\nu}$.
For a test body T 
moving in the gravitational field of the Earth 
as the source S, 
the presence of nonzero $\ol a^w_\mu$ 
induces an additional contribution $\widetilde F_z$
to the usual vertical component $F_z$ 
of the laboratory gravitational force in Newton's second law:
\beq
\widetilde F_z = 
- 2 g (\al \oaut_t + \al \oaus_t \mt/\ms).
\label{F}
\eeq
Here, 
$\mt$ and $\ms$ are the masses of T and S,
while $\oaut_t$ and $\oaus_t$ are 
the time components 
of effective coefficients for Lorentz violation for T and S
in the frame $L$.
For a macroscopic test body T 
containing $N^{\rm T}_w$ particles of species $w$
and negligible binding energy,
the effective coefficient for Lorentz violation is
$\ol a^{\rm T}_\mu = \sum_w N^{\rm T}_w \ol a^w_\mu$. 
Similarly, 
the effective coefficient for S 
with $N^{\rm S}_w$ particles of species $w$ is 
$\ol a^{\rm S}_\mu = \sum_w N^{\rm S}_w \ol a^w_\mu$.
Values of 
$N^{\rm T}_w$ can be computed exactly for atoms
and well approximated for laboratory test bodies,
while for $N^{\rm S}_w$ 
recent studies of the bulk Earth composition 
\cite{earth}
yield the estimates 
$N^{\rm S}_e = N^{\rm S}_p 
\simeq N^{\rm S}_n = 1.8 \times 10^{51}$.
Note that $\oaut_t$, $\oaus_t$ are time dependent
because the components $\ol a^w_T$, $\ol a^w_J$
of $\ol a^w_\mu$ are constant in the frame $S$,
and hence the rotation and the revolution of the Earth
induces sidereal and annual time dependences in the component
$\ol a^w_t$ in the frame $L$.

The observable effects from nonzero $\widetilde F_z$ 
are of two basic kinds. 
One arises from the flavor dependence of $\oaut_t$
and hence of $\widetilde F_z$.
This would produce a signal in experiments
testing the weak equivalence principle (WEP),
which compare the gravitational accelerations of two test bodies.
The other effect arises from the time dependence of 
the laboratory-frame components 
$\oaut_t$ and $\oaus_t$
and hence of $\widetilde F_z$.
It would produce a signal 
in gravimeter or other experiments searching for time variations 
in the Newton gravitational coupling $G_N$. 
The transformation between the frames $S$ and $L$
expresses $\ol a^w_t$ in terms of $\ol a^w_T$
and the product of $\ol a^w_J$ and the relevant boost,
which is about $10^{-4}$ for the Earth's revolution 
and about $10^{-6}$ for its rotation.
It follows that WEP tests can achieve sensitivity
to all components $\ol a^w_T$ 
with instantaneous signals and also 
to all components $\ol a^w_J$
with signals involving sidereal or annual variations,
with the latter suppressed by the boost factor.
In contrast,
the single-flavor gravimeter tests
are insensitive to $\ol a^w_T$,
which in this context causes an effect 
equivalent to an unobservable constant rescaling of $G_N$,
but they have boost-suppressed sensitivity 
to the spatial components $\ol a^w_J$
via annual and sidereal variations.

Comparatively few experiments sensitive to $\ol a^w_\mu$ exist,
and so large values of $\ol a^w_\mu$ 
could have remained undetected to date
even for generic models.
If attention is restricted to the constituents of ordinary matter,
up to 12 measurements are needed to constrain
the 12 components $\ol a^w_\mu$ 
($w = e,p,n$; $\mu = T,X,Y,Z$). 
One constraint on the time components $\ol a^w_T$
can be deduced from published data from WEP tests
using a torsion pendulum
with beryllium and titanium test masses
\cite{schlamminger}.
For this experiment,
calculating with Eq.\ \rf{F} yields the constraint 
\bea
| \al \ol a^e_T + \al \ol a^p_T - 0.8\al \ol a^n_T |
< 1 \times 10^{-11} {\rm ~GeV}
\eea
in natural units ($c = \hbar = 1$) at the 90\% confidence level,
where a generic scenario without cancellations is adopted.
Somewhat weaker constraints on similar combinations of coefficients
are implied at order $10^{-8}$ GeV
by older data from WEP tests
with falling corner cubes
\cite{corner}
and at order $10^{-5}$ GeV
by data from WEP tests with atom interferometers
\cite{fray}. 
However,
these constraints can be evaded or suppressed in specific models.
For example,
if the coefficients $\ol a^w_\mu$ are proportional
to the charge $Q$, 
no constraints exist because the effects cancel in neutral matter.
If instead the coefficients are proportional to 
baryon number $B$ or to the mass $m_w$,
then the strongest constraints come from considerations
of the binding energy in the test-body atoms,
and these are weaker than the generic case 
by about an order of magnitude
\cite{kt}.

\begin{figure*}
\renewcommand{\arraystretch}{1.2}
\begin{tabular*}{\textwidth}{@{\extracolsep{\fill}}c|cccc}
\hline
Experiment\ & 
$\al \ol a^w_T$, actual & 
$\al \ol a^w_J$, actual & 
$\al \ol a^w_J$, feasible & 
$\al \ol a^w_T$, future \\
\hline
torsion pendulum 
\cite{schlamminger} &
$10^{-11}$ GeV &
- &
[$10^{-7}$ GeV] &
- \\
falling corner cube 
\cite{corner} &
$10^{-8}$ GeV &
- &
[$10^{-4}$ GeV] &
- \\
atom interferometry 
\cite{fray,peters,dimopoulos} &
$10^{-5}$ GeV &
- &
[$10^{-5}$ GeV] &
\{$10^{-15}$ GeV\} \\
superconducting gravimeter
\cite{warburton} &
- &
- &
[$10^{-6}$ GeV] &
- \\
lunar laser ranging 
\cite{llr} &
- &
- &
[$10^{-6}$ GeV] &
- \\
drop tower 
\cite{bremen} &
- &
- &
- &
\{$10^{-10}$ GeV\} \\
balloon drop 
\cite{balloon} &
- &
- &
- &
\{$10^{-13}$ GeV\} \\
bouncing masses 
\cite{poem} &
- &
- &
- &
\{$10^{-14}$ GeV\} \\
space-based WEP
\cite{spacexpt} & 
- &
- &
- &
\{$10^{-13}$- $10^{-16}$ GeV\} \\
\hline
\end{tabular*}
\vskip5pt
\footnotesize
\bf Table 1. \rm
Actual (this work), 
currently feasible (brackets), 
and future attainable (braces) 
estimated experimental sensitivities.
\end{figure*}

In contrast to the time components $\ol a^w_T$,
the space components $\ol a^w_J$ are presently unconstrained.
Certain existing experiments and data could in principle
yield sensitivity to some combinations 
of $\ol a^w_J$ for generic scenarios.
Analysis of sidereal and annual variations
in the acceleration of falling corner cubes
could reach $10^{-2}$ GeV and $10^{-4}$ GeV,
respectively.
Sidereal measurements with matter interferometers  
at established sensitivities
\cite{peters}
could achieve $10^{-5}$ GeV on  
various components $\al \ol a^w_J$
using different atomic species.
Experiments with torsion pendula 
could attain $10^{-7}$ GeV via sidereal variations
and $10^{-6}$ GeV via annual effects,
the latter being weaker due to centrifugal forces.
Sidereal and annual studies 
with existing types of superconducting gravimeters
could reach $10^{-4}$ GeV and $10^{-6}$ GeV
on various combinations 
of $\al \ol a^w_J$,
assuming sensitivities already attained in classic tests 
\cite{warburton}.
Analysis of annual modulations in available lunar laser ranging data
\cite{llr}
could achieve $10^{-6}$ GeV
on some combinations of $\al \ol a^w_J$.

The prospects for improved measurements 
of $\al \ol a^w_\mu$
in future experiments are excellent,
with gains of several orders of magnitude
on the above estimates being plausible. 
For $\al \ol a^w_T$,
anticipated advances in atom interferometry 
\cite{dimopoulos} 
could make $10^{-15}$ GeV attainable.
Estimated sensitivities for free-fall experiments
imply sensitivities 
of $10^{-10}$ GeV
using a drop tower
\cite{bremen},
$10^{-13}$ GeV
via balloon drop tests 
\cite{balloon},
and of $10^{-14}$ GeV
using bouncing masses in the laboratory
\cite{poem}.
Various space-based WEP tests 
are also currently under development
\cite{spacexpt},
with estimated sensitivities to $\al \ol a^w_T$
of $10^{-13}$ GeV
for microSCOPE
\cite{micro},
of $10^{-15}$ GeV
for Galileo Galilei
\cite{gg},
and of $10^{-16}$ GeV
for STEP
\cite{step}.
All these experiments also offer
potential improvements in measurements of $\al \ol a^w_J$.
The existing limits and estimated attainable sensitivities
on $\al \ol a^w_\mu$ are summarized in Table 1.
Since the space components are presently unconstrained
and only one combination of the time components is measured,
there is considerable room 
for experimental investigation.

The relativity violations involving $\ol a_\mu$
discussed in this work are potentially large,
possibly some 30 orders of magnitude greater
than the best existing sensitivities,
while being countershaded from most experimental observations.
However, they may not be unique.
Other coefficients for relativity violations exist
that are unobservable in Minkowski spacetime
but are observable through gravity couplings
\cite{akgrav}.
This offers interesting prospects for the existence
of a realistic model with comparatively large relativity violations,
generating signals that would be detectable
in gravitational experiments
with current or near-future technology.

This work was supported in part
by DOE grant DE-FG02-91ER40661.


\begin{thebibliography}{99}

\bibitem{tables}
{\it Data Tables for Lorentz and CPT Violation,}
V.A.\ Kosteleck\'y and N.\ Russell,
arXiv:0801.0287.

\bibitem{will}
C.M.\ Will,
\it Theory and Experiment in Gravitational Physics, \rm
Cambridge University Press, Cambridge, 1993.

\bibitem{gravexpt}
H.\ M\"uller, S.\ Chiow, S.\ Herrmann, S.\ Chu, and K.-Y.\ Chung,
Phys.\ Rev.\ Lett.\ 
{\bf 100}, 031101 (2008);
J.B.R.\ Battat, J.F.\ Chandler, and C.W.\ Stubbs,
Phys.\ Rev.\ Lett.\ 
{\bf 99}, 241103 (2007);
Q.G.\ Bailey and V.A.\ Kosteleck\'y,
Phys.\ Rev.\ D {\bf 74}, 045001 (2006).

\bibitem{kp95}
V.A.\ Kosteleck\'y and R.\ Potting,
Phys.\ Rev.\ D 
{\bf 51}, 3923 (1995).

\bibitem{ck}
D.\ Colladay and V.A.\ Kosteleck\'y, 
Phys.\ Rev.\ D 
{\bf 55}, 6760 (1997);
Phys.\ Rev.\ D 
{\bf 58}, 116002 (1998).

\bibitem{mesons}
BaBar Collaboration,
B.\ Aubert \etal,
Phys.\ Rev.\ Lett.\ 
{\bf 100}, 131802 (2008);
FOCUS Collaboration,
J.M.\ Link \etal,
Phys.\ Lett.\ B {\bf 556}, 7 (2003);
KLOE Collaboration,
M.\ Testa \etal,
arXiv:0805.1969;
KTEV Collaboration,
H.\ Nguyen \etal,
arXiv:hep-ex/0112046;
V.A.\ Kosteleck\'y,
Phys.\ Rev.\ Lett.\ 
{\bf 80}, 1818 (1998);
Phys.\ Rev.\ D {\bf 61}, 016002 (2000);
Phys.\ Rev.\ D {\bf 64}, 076001 (2001).

\bibitem{nu}
MINOS Collaboration,
P.\ Adamson \etal,
Phys.\ Rev.\ Lett., 
in press
[arXiv:0806.4945];
LSND Collaboration,
L.B.\ Auerbach \etal,
Phys.\ Rev.\ D 
{\bf 72}, 076004 (2005);
V.A.\ Kosteleck\'y and M.\ Mewes,
Phys.\ Rev.\ D {\bf 69}, 016005 (2004);
Phys.\ Rev.\ D {\bf 70}, 031902(R) (2004);
Phys.\ Rev.\ D {\bf 70}, 076002 (2004);
T.\ Katori \etal,
Phys.\ Rev.\ D {\bf 74}, 105009 (2006);
V.\ Barger, D.\ Marfatia, and K.\ Whisnant,
Phys.\ Lett.\ B {\bf 653}, 267 (2007).

\bibitem{akgrav}
V.A.\ Kosteleck\'y,
Phys.\ Rev.\ D 
{\bf 69}, 105009 (2004).

\bibitem{ks}
V.A.\ Kosteleck\'y and S.\ Samuel,
Phys.\ Rev.\ D {\bf 39}, 683 (1989);
Phys.\ Rev.\ D {\bf 40}, 1886 (1989).

\bibitem{kl}
V.A.\ Kosteleck\'y and R.\ Lehnert,
Phys.\ Rev.\ D {\bf 63}, 065008 (2001).

\bibitem{bkgrav}
R.\ Bluhm and V.A.\ Kosteleck\'y,
Phys.\ Rev.\ D 
{\bf 71}, 065008 (2005);
B.\ Altschul and V.A.\ Kosteleck\'y,
Phys.\ Lett.\ B {\bf 628}, 106 (2005).

\bibitem{kpgrav}
V.A.\ Kosteleck\'y and R.\ Potting,
Gen.\ Rel.\ Grav.\ 
{\bf 37}, 1675 (2005).

\bibitem{ah}
N.\ Arkani-Hamed, H.-C.\ Cheng, M.\ Luty and J.\ Thaler,
JHEP 
{\bf 0507}, 029 (2005).

\bibitem{schlamminger}
S.\ Schlamminger, K.-Y.\ Choi, T.A.\ Wagner, 
J.H.\ Gundlach, and E.G.\ Adelberger, 
Phys.\ Rev.\ Lett.\ 
{\bf 100}, 041101 (2008).

\bibitem{emepspace}
H.\ Dittus, C.\ L\"ammerzahl, and H.\ Selig, 
Gen.\ Rel.\ Grav.\ 
{\bf 36}, 571 (2004).

\bibitem{antih}
AEGIS Collaboration,
A.\ Kellerbauer \etal,
Nucl.\ Instrum.\ Meth.\ {\bf B266}, 351 (2008);
J.\ Walz and T.W.\ H\"ansch,
Gen.\ Rel.\ Grav.\ {\bf 36}, 561 (2004);
M.M.\ Nieto and J.T.\ Goldman,
Phys.\ Rep.\ {\bf 205}, 221 (1991).

\bibitem{muonium}
K.\ Kirch,
arXiv:physics/0702143;
M.\ Oberthaler,
Nucl.\ Instrum.\ Meth.\ {\bf B192}, 129 (2002).

\bibitem{km}
V.A.\ Kosteleck\'y and M.\ Mewes,
Phys.\ Rev.\ D 
{\bf 66}, 056005 (2002).

\bibitem{earth}
C.J.\ All\`egre, J.-P.\ Poirier, E.\ Humler, and A.W.\ Hofmann,
Earth Planet.\ Sci.\ Lett.\ 
{\bf 134}, 515-526 (1995).

\bibitem{corner}
K.\ Kuroda and N.\ Mio,
Phys.\ Rev.\ D 
{\bf 42}, 3903 (1990);
T.M.\ Niebauer, M.P.\ McHugh, and J.E.\ Faller,
Phys.\ Rev.\ Lett.\ 
{\bf 59}, 609 (1987).

\bibitem{fray}
S.\ Fray, C.A.\ Diez, T.W.\ H\"ansch, and M.\ Weitz,
Phys.\ Rev.\ Lett.\
{\bf 93}, 240404 (2004).

\bibitem{peters}
A.\ Peters, K.Y.\ Chung, and S.\ Chu,
Nature 
{\bf 400}, 849 (1999);
Metrologia {\bf 38}, 25 (2001).

\bibitem{kt}
V.A.\ Kosteleck\'y and J.D.\ Tasson,
in preparation.
 
\bibitem{warburton}
R.J.\ Warburton and J.M.\ Goodkind,
Astrophys.\ J.\ 
{\bf 208}, 881 (1976).

\bibitem{llr}
J.G.\ Williams, S.G.\ Turyshev and H.D. Boggs, 
Phys.\ Rev.\ Lett.\ 
{\bf 93}, 261101 (2004).

\bibitem{dimopoulos}
S.\ Dimopoulos, P.W.\ Graham, J.M.\ Hogan, and M.A.\ Kasevich,
Phys.\ Rev.\ Lett.\ 
{\bf 98}, 111102 (2007).

\bibitem{bremen}
H.\ Dittus and C.\ Mehls,
Class.\ Quant.\ Grav.\ 
{\bf 18}, 2417 (2001).

\bibitem{balloon}
V.\ Iafolla, S.\ Nozzoli, E.C.\ Lorenzini, 
I.I.\ Shapiro, and V.\ Milyukov,
Class.\ Quant.\ Grav.\ 
{\bf 17}, 2327 (2000).

\bibitem{poem}
R.\ Reasenberg and J.D.\ Phillips,
in V.A.\ Kosteleck\'y, ed.,
{\it CPT and Lorentz Symmetry IV},
World Scientific, Singapore, 2008.

\bibitem{spacexpt}
C.\ L\"ammerzahl, C.W.F.\ Everitt, and F.W.\ Hehl, eds.,
{ \it Gyros, Clocks, interferometers\ldots :
Testing Relativistic Gravity in Space},
Springer, Berlin, 2001.

\bibitem{micro}
P.\ Touboul,
in Ref.\ \cite{spacexpt}, p.\ 273.

\bibitem{gg}
A.M.\ Nobili \etal,
Phys.\ Lett.\ A {\bf 318}, 172 (2003).

\bibitem{step}
N.\ Lockerbie, J.C.\ Mester, R.\ Torii, 
S.\ Vitale, and P.W.\ Worden, 
in Ref.\ \cite{spacexpt}, p. 213.

\end{thebibliography}
\end{document}